\documentclass[aps, twocolumn, groupedaddress,superscriptaddress, longbibliography, showpacs,prx,nofootinbib]{revtex4-2}
\usepackage[utf8]{inputenc}
\usepackage[T1]{fontenc}
\usepackage{amsmath}
\usepackage{amsfonts}
\usepackage{amssymb}
\usepackage{graphicx}
\usepackage{lineno}
\usepackage{xcolor}
\usepackage{hyperref}

\begin{document}

\title{Dynamics of critical cascades in interdependent networks}

\author{Dolev Dilmoney}
\affiliation{Department of Physics, Bar-Ilan University, Ramat-Gan IL52900, Israel.}
\author{Bnaya Gross}
\affiliation{Network Science Institute, Northeastern University, Boston, MA
02115}
\affiliation{Department of Physics, Northeastern University, Boston, MA 02115}
\author{Shlomo Havlin}
\affiliation{Department of Physics, Bar-Ilan University, Ramat-Gan IL52900, Israel.}
\author{Nadav M. Shnerb}
\affiliation{Department of Physics, Bar-Ilan University, Ramat-Gan IL52900, Israel.}

\begin{abstract}
\noindent The collapse of interdependent networks, as well as similar avalanche phenomena, is driven by cascading failures. At the critical point, the cascade begins as a critical branching process, where each failing node (element) triggers, on average, the failure of one other node. As nodes continue to fail, the network becomes increasingly fragile and the branching factor grows. If the failure process does not reach extinction during its critical phase, the network undergoes an abrupt collapse. Here, we implement the analogy between this dynamic and birth-death processes to derive new analytical results and significantly optimize numerical calculations. Using this approach, we analyze three key aspects of the dynamics: the probability of collapse, the duration of avalanches, and the length of the cascading plateau phase preceding a collapse. This analysis quantifies how system size and the intensity of the initial triggering event influence these characteristics.
\end{abstract}
\maketitle

Network science has experienced remarkable development in recent decades, finding extensive applications across various fields, including the study of neural, social, metabolic, and communication networks~\cite{barabasi2009scale,mata2020complex}. Within this domain, the topic of interdependent networks has gained significant attention in recent years~\cite{buldyrev2010catastrophic}. This concept has also found a wide range of applications, from critical infrastructures and cybersecurity to the dynamics of ecological communities~\cite{pocock2012robustness,ouyang2014review,rocha2018cascading}. Recent work has effectively demonstrated the key features of interdependent network collapse using a clear and physically straightforward experimental model, which involves thermally connected layers of amorphous superconductors~\cite{bonamassa2023interdependent}.

\begin{figure*}[hbt!]
	\centering{
    	\includegraphics[width=0.9\textwidth]{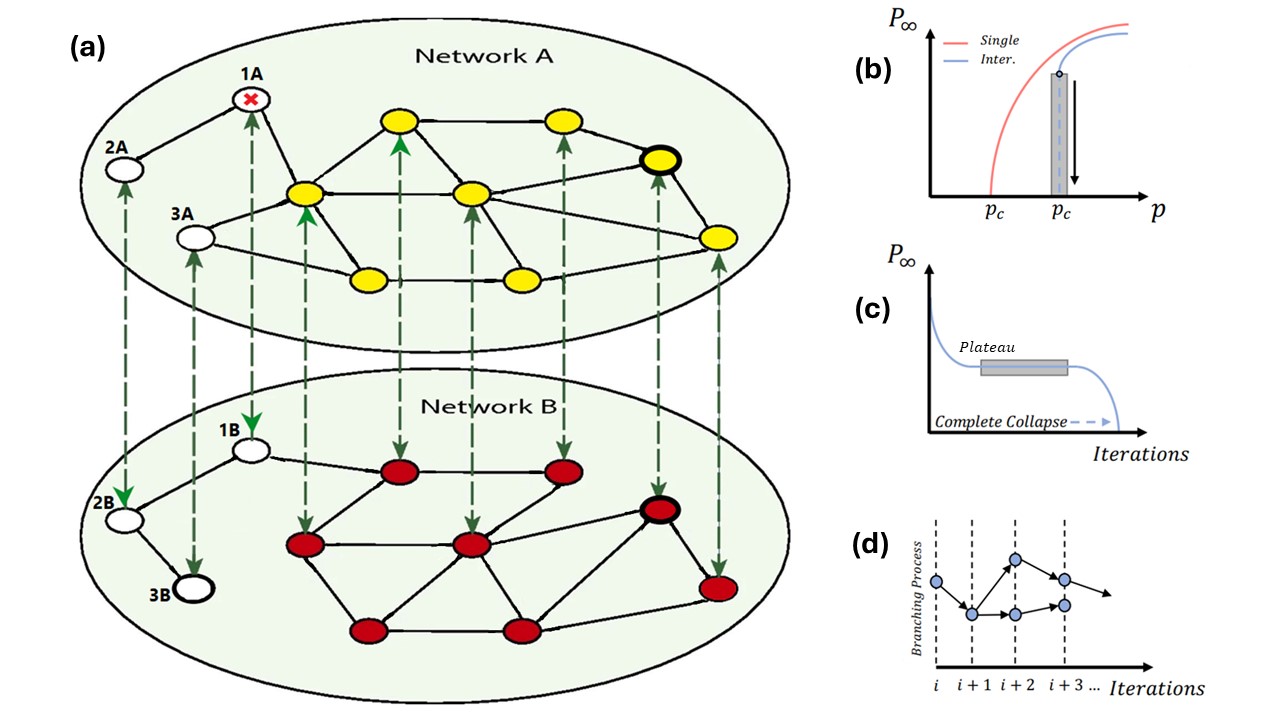}}
	\caption{ \textbf{Cascading failures in interdependent networks conceptualized as a birth-death process.} Panel \textbf{(a)} depicts two networks, A and B, interconnected by dependency links (green) alongside internal connectivity links (black). When a node fails, it topples down (in the same network) all the nodes it disconnects from GCC. Additionally, each failed node also causes the dependent node in the parallel network to fail, leading to a cascading effect. For instance, the failure of node 1A (marked with red X) results in the detachment of 2A from the GCC, causing the failure of nodes 1B and 2B due to dependency. Consequently, 3B becomes detached from the GCC of network B, which cascades back to topple 3A. Thus the failure of 1A leads to the failure of 3A in the subsequent iteration.  Panel \textbf{(b)}: while single networks exhibit a continuous phase transition (red) in which the likelihood of a node being part of the GCC ($P_\infty$) continuously diminishes to zero, interdependent networks undergo an abrupt transition (blue) due to cascading failures. At $p={\tilde p}$ (Panel \textbf{(c)}), the cascade of failures involves a long-lasting plateau, where damage stays relatively constant over several iterations. Panel \textbf{(d)}:  The cascading failure dynamics mirrors a birth-death process, wherein each failed node [like 1A in panel (a)] has a given number of offspring (like 3A) in the subsequent iteration. When $p>{\tilde p}$ the mean offspring count, ${\overline m}$, is less than one, resulting in an exponential decay of the demographic process and network survival. Conversely, for $p<{\tilde p}$, ${\overline m}>1$ leading to exponential damage growth and a swift collapse. At $p={\tilde p}$, ${\overline m}=1$ during the plateau, and the dynamics is neutral}
 \label{fig1}
\end{figure*}

Interdependent networks are characterized by two types of links: \textit{connectivity} links within each network and \textit{dependency} links between nodes of different networks. Each link type serves a distinct role. Connectivity links are understood as in standard percolation problems, so nodes that disconnect from the giant connected component (GCC) of their network are considered non-functional. Dependency links, on the other hand, propagate damage between the networks: if a node in one network fails, its dependent node in the other network will fail as well, even if it is still connected to the GCC of its own network (see Figure~\ref{fig1}\textbf{(a)}). When nodes are removed from one network, the interplay between these two types of links can trigger a cascading failure. Therefore, while a single network experiences a continuous second-order transition in which the GCC collapses when $1-p_c$ of the nodes are removed, interdependent networks undergo an abrupt, first order transition, at ${\tilde p}>p_c$ (Figure 1({\bf b}))~\cite{buldyrev2010catastrophic}. 

These cascading failures are a general characteristic of avalanche phenomena~\cite{lee2016hybrid}, which have also been extensively studied in recent decades in various contexts such as magnetic materials, plastic distortions, crack propagation, earthquakes, neural networks and so on~\cite{perkovic1995avalanches,sethna2001crackling,haldeman2005critical,friedman2012universal,gross2024microscopic}. In all these systems, a failure or change in the state of a single component, such as a spin, imposes 'pressure' on adjacent components, and a critical point in parameter space separates localized avalanches from widespread collapses. Such cascades are mathematically analogous to stochastic birth-death processes, where individuals can produce a random number of offspring~\cite{harris1963theory,lomnitz1992avalanches,haldeman2005critical,shekhawat2013damage,zhou2014simultaneous}. 

In this Letter, we aim to examine the analogy between demographic processes and the dynamics of cascading failures. In particular, we will focus on the critical point, where the corresponding dynamics begins as a neutral birth-death process~\cite{kimura1985role,hubbell_book,azaele2015towards,azaele2016statistical} but, over time, diverges from neutrality. We will implement this analogy to derive novel qualitative and quantitative insights.

Specifically, our study addresses three crucial questions: First, how does the likelihood of a collapse change in relation to the scale of the initial attack? Second, the impact of this initial attack size on the 'plateau time' — the period of slow dynamics preceding the collapse (Figure ~\ref{fig1}\textbf{(c)}).  Third, we examine how the initial attack size impacts the mean 'avalanche time,' the duration until an avalanche ends, whether or not it causes collapse.

\underline{\textit{From Nodes to Generations:--}} As depicted in Figure 1({\textbf a}), damage cascades reciprocally between dependent networks, with each failed node having a chance $P_{1 \to m}$ of triggering the failure of $m$ nodes within the same network in the subsequent iteration. With a sufficiently large number of nodes, $N$, and a small number of failed nodes, failures are uncorrelated. Consequently, at the $t$-th step, if the count of failed nodes is $n_t$, then $n_{t+1}$—the tally of failed nodes in the subsequent step in the same network—is the sum of $n_t$ independent variables,  following a probability distribution function of $P_{1 \to m}$ (see Appendix  \ref{apA}). Hence, this branching process mirrors a demographic dynamics, where nodes represent individuals and iterations are equivalent to generations (Figure 1({\bf d})). At $p={\tilde p}$, the process is {\it neutral}, meaning that during the plateau (Figure 1({\bf c})), the mean number of offspring per node is one. 

The ``population dynamics'' of $n_t$ is influenced by demographic stochasticity, namely the binomial noise associated with the randomness of the birth-death process, and to deterministic forces that yield exponential growth (if  $\overline{m}>1$) or exponential decay  (if $\overline{m}<1$). At ${\tilde p}$ the growth rate vanishes, and the dynamics is neutral:  the number of failures in each generation, $n$,  is (on average) time-independent, and its time evolution is governed solely by demographic noise. Because $n \ll N$, the GCC undergoes a prolonged plateau, in which $P_\infty$ is almost fixed (Figure 1({\bf c})). This plateau phase may end in extinction, in which case  $n_t=0$ and the network survives.  However, if the process persists long enough, the GCC becomes thinner and more fragile,  $\overline{m}$  increases above one, the population of failed nodes starts growing faster and the network collapses.

\begin{figure*}[hbt!]
	\centering{
		\includegraphics[width=0.45\textwidth]{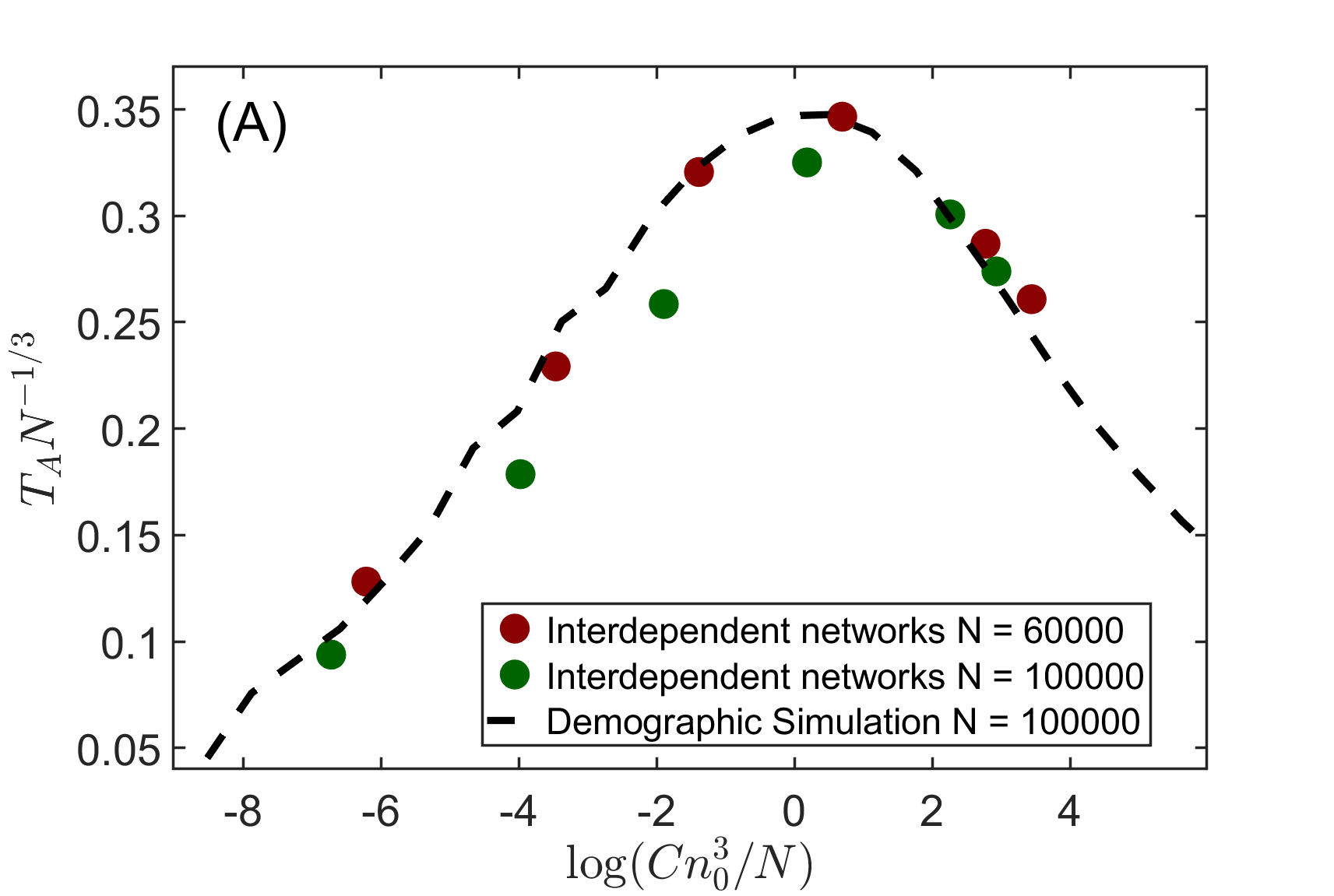}
\includegraphics[width=0.45\textwidth]{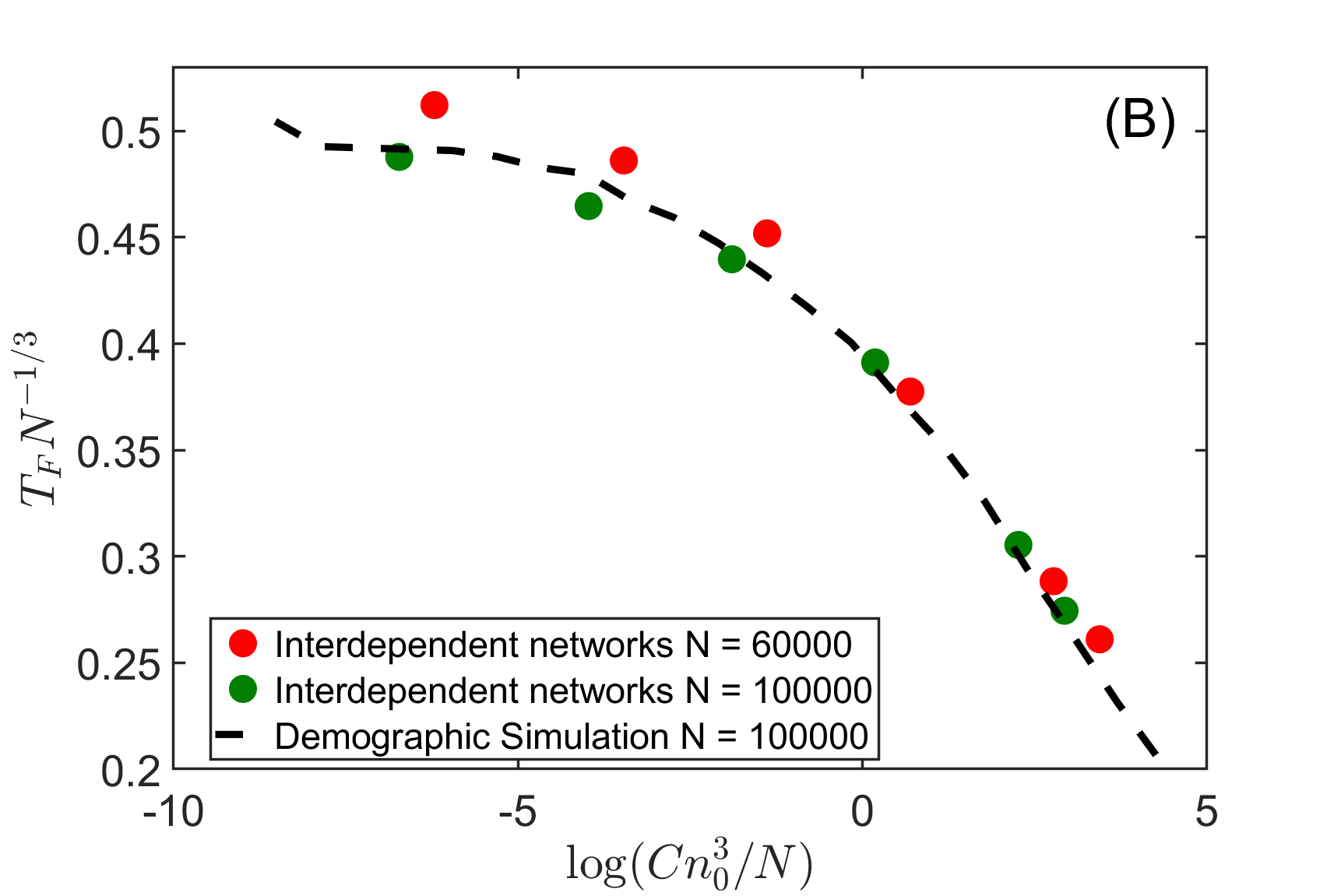}}
	\caption{{\bf  A comparison between the dynamics of interdependent networks and of the correspondent demographic process.} Panel \textbf{(a)} shows the mean time to absorption $T_A$ (time until the cascading failure process ends, either in a collapse of the network or in its survival) and Panel \textbf{(b)}, the mean time to failure $T_F$ (time until the cascading process ends in a failure, conditioned on failure). In both panels, markers present the outcome of an explicit interdependent networks dynamics, whereas the dashed lines show the results of the corresponding demographic process. We used Erd\"{o}s–R\'{e}nyi networks with average degree $k=5$, and the results present an average over $10^4$ different network realizations for each $N$. Note that the larger the network, the better the fit to the simulations of demographic theory. For additional details and metrics regarding the distribution of outcomes, see Appendix  \ref{appE}.  The basic scaling with $N^{1/3}$, implied in these graphs, is derived below following Eq. (\ref{eq7}). \label{fig3}}
\end{figure*}

 Let us provide an approximate expression for the deviation from neutrality as the process goes on. $t$ is defined as the number of iterations since the beginning of the plateau, so at $t=0$  the chance to be on the infinite cluster is $P_{\infty}(t=0)$, the number of failed nodes is $n_0$ and the mean number of offspring ${\overline{m}}(t=0)=1$. The number of nodes that fail in the $t$-th iteration is $n_t$, and $M_t = \sum_{k=0}^{t} n_t$ is the total number of nodes that fail until $t$.  Therefore, $P_\infty(t) =P_\infty(0) -M_t/N$. As long as $M_t \ll N$ we can implement the linear approximation and assume ${\overline m}(t) = 1+C M_t/N$, where $C \equiv d{\overline m}/dP_\infty$ evaluated at $P_\infty(t=0)$.

To simulate such a demographic process, we start with initial values \( n_0 = M_0 \). For each individual in \( n_0 \), we randomly draw a number \( m_i \) of descendants for the \( i \)th individual from a truncated power-law distribution with mean \( 1 + CM/N \) (see Appendix~\ref{apR} for full details). In the next iteration, the population is \( n_1 \equiv \sum_{i=1}^{n_0} m_i \). We then update \( M_0 \) to \( M_1 = M_0 + n_1 \) and iterate the process until it either terminates (i.e., when \( n_t = 0 \)) or the network collapses, which can be approximated by the condition \( M_t = (\tilde{p} - p_c)N \). Such simulations are significantly simpler and faster than emulating dependent networks and their dynamics. Monitoring the giant component of a random network is computationally costly, requiring the tracking of all node and edge connections at each step. In contrast, simulating the dynamics of a neutral population only involves drawing \( n \) random deviates at each step.

In order to demonstrate the  effectiveness of our approach, we will use it to understand how the collapse dynamics depend on the strength of the initial attack, that is, on the number of nodes that failed in the first step, $n_0$. Any such attack can end either when the GCC in both networks still exists, in which case we say that the attack has failed.  The attack is considered successful if it leads to a collapse. 

In general, the analysis of such processes involves three fundamental questions.  First, what is the probability of success, second, what is the average number of iterations until success, assuming the attack ended successfully, and third, what is the average number of iterations until the attack is over, where in this third case we do not distinguish between successful and failed attacks. At the critical point $p={\tilde p}$, all three of these quantities depend on the size of the network $N$ and on the strength of the initial attack $n_0$. As will be shown later, the dependence on $n_0$ enters through the variable $n_0^3/N$.

Let's start by comparing two types of numerical results: those obtained from a direct simulation of interdependent networks, a complex simulation that requires building a suitable network and tracking the size of the mutual GCC at each step, versus the simple and fast simulation of the corresponding demographic process.

For direct simulations, we prepared two interdependent networks at their critical state. At the critical point, the dynamics within each network is such that, on average, each failing node triggers the detachment of {\it one} other node from the GCC. Accordingly, we identified the corresponding value of $p_c$ by analyzing a single network. Then we duplicate it and connect the nodes of the two resulting networks via randomly chosen dependency links. See Appendix  \ref{apD} for further details.

Once we have a system at criticality, we can examine the characteristics of the collapse process and compare it with the results obtained for the demographic process as described above.

Figure \ref{fig3} demonstrates the power of our technique. For the same $n_0$, we examined the mean time until collapse $T_F$, and the mean time to absorption $T_A$, and compare the demographic process (where time is measured in generations) with the collapse of the interdependent networks (where time is measured in iterations). The qualitative and quantitative match is very good, which demonstrates that the demographic model indeed reflects the dynamics of the networks.

After establishing a quantitative correspondence between the two processes, we can apply well-known techniques from the neutral theory~\cite{kimura1985role,karlin2014first} to derive analytical solutions for interdependent networks problems. To begin, let us consider $\Pi(n_0,N)$, the probability of collapse given $n_0$ and $N$. The analogous problem in neutral dynamics is the probability of ultimate fixation.

Switching to continuous time, the Langevin equation for a population under pure demographic stochasticity   is  $dn/dt = \eta(t)\sqrt{n}$, where $\eta(t)$ is a white noise process~\cite{pechenik1999interfacial}.
As explained, at criticality the cascading failure process corresponds to the dynamics of a population for which the growth rate is $CM(t)/N$, where $M(t)$ is the accumulated number of birth (failed nodes) until $t$ and $C$ is a constant.  Accordingly, the coupled Langevin equations characterizing the interdependent network process are delineated as follows:
\begin{equation} \label{eq1}
\frac{dn(t)}{dt} = \frac{C M(t)}{N}n + \eta(t)\sqrt{n}, \qquad \frac{dM(t)}{dt} = n.
\end{equation}
 As long as the process is neutral (i.e., when $M/N$ is small),  a process that survived until $t$ admits $n(t)  \sim t$ (See Appendix  \ref{apB} for further details) and $M(t) \sim t^2$.  Therefore, $M \sim n^2$, and hence  Eq.  (\ref{eq1}) may be approximated as,
\begin{equation} \label{eq2}
\frac{dn(t)}{dt} = C n^3/N + \eta(t)\sqrt{n}.
\end{equation}
The corresponding  backward Kolmogorov equation for the chance of network collapse~\cite{karlin2014first}, $\Pi$, as a function of the initial damage $n_0$, can be written for the process described in Eq. (\ref{eq2}) as,
    \begin{equation} \label{eq4}
n \Pi''(n) + C \frac{n^3}{N} \Pi'(n) =0, \qquad \Pi(0)=0, \qquad \Pi(N^*)=1,
\end{equation}
where $N^* \equiv N({\tilde p}-p_c)$.
The solution is
\begin{equation} \label{eq5}
\Pi(n_0) = 1-\Gamma(1/3,z/3)/\Gamma(1/3),
\end{equation}
where $z =C n_0^3/N^*$ and $\Gamma(a,x) = \int_x^\infty t^{a-1} \exp(-t) dt$.

\begin{figure}[hbt!]
	\centering{
		\includegraphics[width=0.4\textwidth]{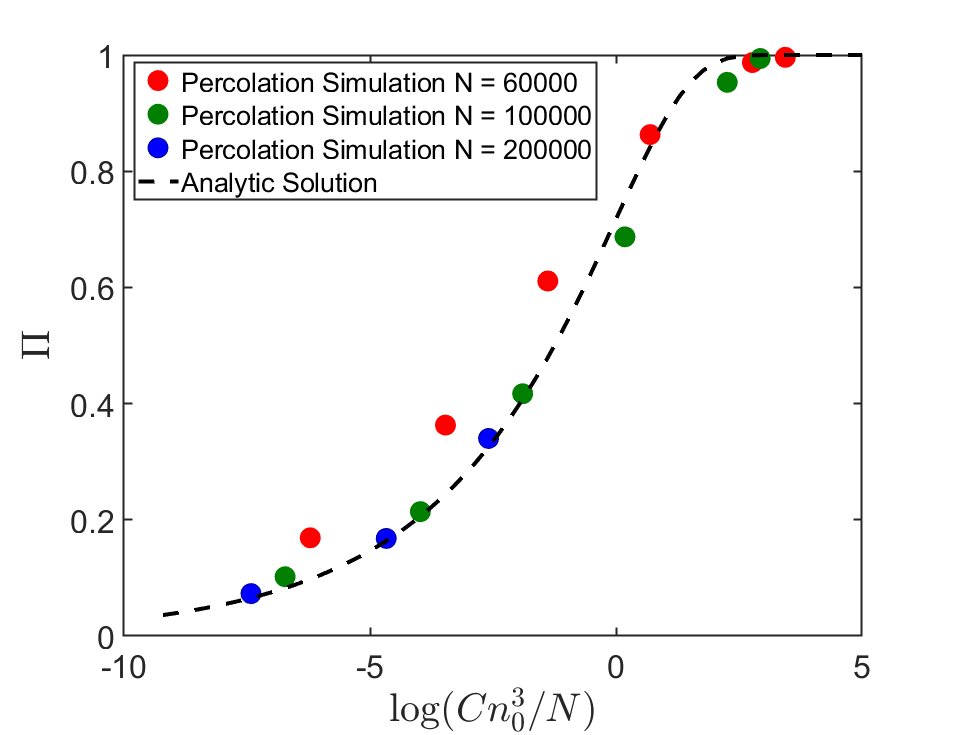}}
	\caption{\textbf{The chance of collapse, $\Pi$, as a function of the initial damage.} The dashed line shows $\Pi(z)$ as a function of $z = C n_0^3/N$, as predicted in Eq. (\ref{eq5}), for the chance of a demographic process that satisfies Eq.~(\ref{eq2}) to reach fixation at $n=N$. Each circle corresponds to the chance of collapse of interdependent networks for a given $N$ (as indicated in the legend) and $n_0$, as obtained from an average over $10^3$ random initial conditions.     \label{fig4}}
\end{figure}

In Figure \ref{fig4} we see a comparison between this solution and the chance for two interdependent networks will collapse, as obtained from direct simulations.

Another immediate result of Eq. (\ref{eq2}) is the general scaling behavior at criticality.  The Langevin equation for a stochastic population  and selection $s$   is given  by~\cite{karlin2014first},
\begin{equation} \label{eq7}
\frac{dn(t)}{dt} = s n + \eta(t)\sqrt{n}.
\end{equation}

 Comparing  Eq. (\ref{eq7}) with  Eq. (\ref{eq2}) one can see that $s$ is analogous to $C n^2/N$.
In processes described by Eq. (\ref{eq7}), the stochastic component becomes negligible (i.e., the process becomes, effectively, deterministic exponential growth) at $n \approx 1/s$~\cite{desai2007speed}.  Equivalently, the transition for interdependent networks takes place at  $n \approx N/C n^2$. Therefore, the basic scaling of the cascading collapse process is $n^3 \sim T$; since under neutral dynamics $n \sim T$, we obtain $T \sim N^{1/3}$, in agreement with the simulation results of~\cite{zhou2014simultaneous}.

\underline{\textit{Discussion:--}} 
In large systems, cascading failure processes can be described as branching processes~\cite{harris1963theory}, where each failed element has a certain probability of causing additional elements to fail. This situation is analogous to a demographic process in which each individual has a probability of producing a certain number of offspring. The process becomes critical, or "neutral," when the average number of offspring per element equals one.

In a neutral process, the probability of an individual lineage surviving indefinitely approaches zero over infinite time. Since individuals are uncorrelated, for any finite number of initially failed elements, the cascading failure must eventually stop if the system is large enough. However, in the case of interdependent networks, the system becomes increasingly fragile as the number of failed components grows. Therefore, if the process persists long enough, it will become supercritical, and the network will collapse.

The key question is how fragility increases relative to the extent of the damage. In our case, the deviation from criticality is determined by the ratio between the number of failed components and the system size. Consequently, the time during which the process remains approximately neutral depends on $N$ and, in our case, scales as $N^{1/3}$. If the damage requires for a substantial deviation from criticality does not depend on $N$, one observes a standard nucleation process, where the amount of damage required to ensure transition to the propagative phase does not scale with the system size.


It is noteworthy that, overall, our avalanche process resembles the spread of epidemics at their critical point~\cite{ben2004size,kessler2007solution,kessler2008epidemic}. In both cases, as long as the total number of infected individuals remains much smaller than the population size $N$, the process is entirely neutral. Furthermore, as in our case, when the demographic process persists long enough and the number of infected individuals scales with $N$ to some power, the dynamics become non-neutral. The key difference between avalanches and epidemics lies in the \emph{sign} of the effect. In ecological demographic processes, the number of offspring per individual typically {\it decreases} as population size increases, leading to a continuous, second-order transition. For example, in infection processes, as more individuals become infected, fewer remain susceptible. In contrast, in interdependent networks, the opposite effect occurs: as more nodes fail,  the number of offspring (failures) increases, resulting in a discontinuous, first-order collapse. The relevant equation (Eq. \ref{eq2}) describes a process that reaches an infinite population in finite time, so the central question is whether the neutral process can survive long enough for $M(t)$ to reach a level where significant deterministic growth occurs.


This type of dynamics, where the failure of one component triggers a cascading failure, is not unique to interdependent networks. It appears in many avalanche-like processes, such as fracture dynamics, magnetic systems, and socioecological transitions~\cite{shekhawat2013damage,perkovic1995avalanches,sethna2001crackling,hagstrom2023phase}. In all these systems, criticality is defined by each failing element, on average, triggering the failure of one additional element. Moreover, while our analysis focuses on Erdős–Rényi networks with low clustering, this defining criterion suggests that the same critical behavior should emerge in other network types as well. In other network architectures (e.g., scale-free networks) the variance in the number of triggered failures may be higher, potentially making them more robust. Testing this hypothesis, however, requires further study.

{\bf Acknowledgments:} N.M.S acknowledges support from Israel Science Foundation (grant no. 2435/24). B.G. acknowledges the support of the Fulbright Postdoctoral Fellowship Program. S.H. thanks the support of the Israel Science Foundation (Grant No. 189/19), the EU  H2020 DIT4Tram (Grant number 953783) and the Horizon Europe grant
2 OMINO (Grant number 101086321).

\bibliography{mybib}

\clearpage
\onecolumngrid
\appendix

\section{The demographic process} \label{apA}

As explained in the main text, the demographic process in interdependent networks involves three steps within each "generation".

\begin{itemize}
  \item First, an internetwork failure due to detachment: each node failing in network A causes the detachment of $\ell$ other nodes from the GCC of A, with probability $P_\ell$.
  \item Second, due to dependency, the events in A lead to $\ell+1$ failures of nodes in network B. The failure of these nodes does not cascade back, since their dependent nodes have already collapsed.
  \item What cascade back is the effect of the $m$ extra nodes that are detached from the GCC of B due to these $\ell +1$ failures.
\end{itemize}

   Therefore, if ${ m_1, m_2, ... m_{\ell+1} }$ are the number of B nodes that detach from the GCC due to the failure of nodes $1..\ell+1$, the total number of "descendants" of a single node in A is $m=\sum_{i=1,\ell+1} m_i$. In other words, the number of offspring  satisfies:
\begin{equation} \label{eqA1}
P_{1 \to m} = \sum_{\ell=0}^{\infty} P^A_{1 \to \ell} P^B_{\ell+1 \to m},
\end{equation}
where $P^A$ and $P^B$ denote the detachment probabilities in networks A and B, respectively. Throughout this paper, we assume that the two networks are statistically identical, thus $P^A=P^B$. In Eq. (\ref{eqA1}), $P_{1 \to m}$ (without a superscript) represents the probability of $m$ being the overall number of offspring per individual, that is, $m$ is the number of nodes in network A, for instance, that fail due to the back reaction from network B in the next iteration of the cascading failure.

\section{Preparing interdependent networks at their critical state} \label{apD}

 In this paper, we report results such as the probability of failure as a function of the initial number of failing nodes, for two interdependent networks at criticality. In this section, we explain the setup of the numerical experiment, specifically how to prepare two interdependent networks at their critical state.

We first considered a {\it single} network, A, and attacked it with a certain probability, $p$. Following the attack, we randomly removed one node from the GCC and recorded $\ell$, the number of nodes that subsequently became disconnected from the GCC due to this removal (see Appendix \ref{apA}). By repeating this process multiple times, we obtained a reliable average for the value of $\ell$. We then used a bisection search method to identify the $p$-value where ${\overline \ell}$ is approximately equal to one. For an ER network with a mean degree of $k=5$, this occurs around $p=0.35$.

Once the value of $p$ for which ${\overline \ell}=1$ was identified, network B was created by duplicating network A. Each node in network A was then  connected to a (randomly picked) node in network B via a dependency link, thereby forming a pair of interdependent networks at their critical point.

\section{Neutral dynamics} \label{apB}

In the main text, we utilized several features of the neutral process, particularly the probability of the process surviving until time $t$ and the population size, $n_t$, conditioned on survival. These results have been derived by many authors (see, e.g., \cite{pechenik1999interfacial}). Here, we provide a brief summary of the relevant formulas.

First, the chance of the lineage of a single individual to survive for $t$ generations. For a population of $n$ independent individuals, each having birth rate  $b$ and  death rate  $d$, the corresponding master equation is,
\begin{equation} \label{eqa1}
    \frac{dP_n}{dt} =  -(b+d)n P_n + b(n-1)P_{n-1} + d(n+1)P_{n+1},
\end{equation}
where $P_n(t)$ is the chance to find $n$ individuals at time $t$. The dynamics is neutral if $b=d$, and a natural choice of a generation time is $b=d=1$. In such a case, the distribution of the total number of offspring per individual (i.e., the number of offspring it generates before its death) is geometric, $P_{1 \to \ell} = 1/2^{\ell+1}$, with mean one and variance is $\sigma^2 =2$. 

This process is analytically solvable (e.g., using characteristics). The chance that the lineage goes extinct before the $t$-th generation is 
\begin{equation} \label{Eqap1}
Q(t) = \frac{t}{1+t} \approx 1-\frac{1}{t},
\end{equation}
where the last approximation is for large $t$. Thus, the probability of survival for $t$ generations is 
\begin{equation}
S(t) = 1-Q(t) \approx \frac{1}{t}.
\end{equation}

Conditioned on survival, the chance that the population has size $n$ at time $t$ is 
\begin{equation}
P_n(t) = \frac{1}{t} \left(\frac{t}{1+t}\right)^n \approx \frac{e^{-n/t}}{t}
\end{equation}
and therefore, the mean population size (conditioned on survival) is ${\overline n(t)} =t$. 

 Since ${\overline n(t)} =t$, the  accumulated number of birth until $t$, $M(t)$, is proportional (again, for surviving populations) to $t^2$. 

\begin{figure}[hbt!]
	\centering{
		\includegraphics[width=0.4\textwidth]{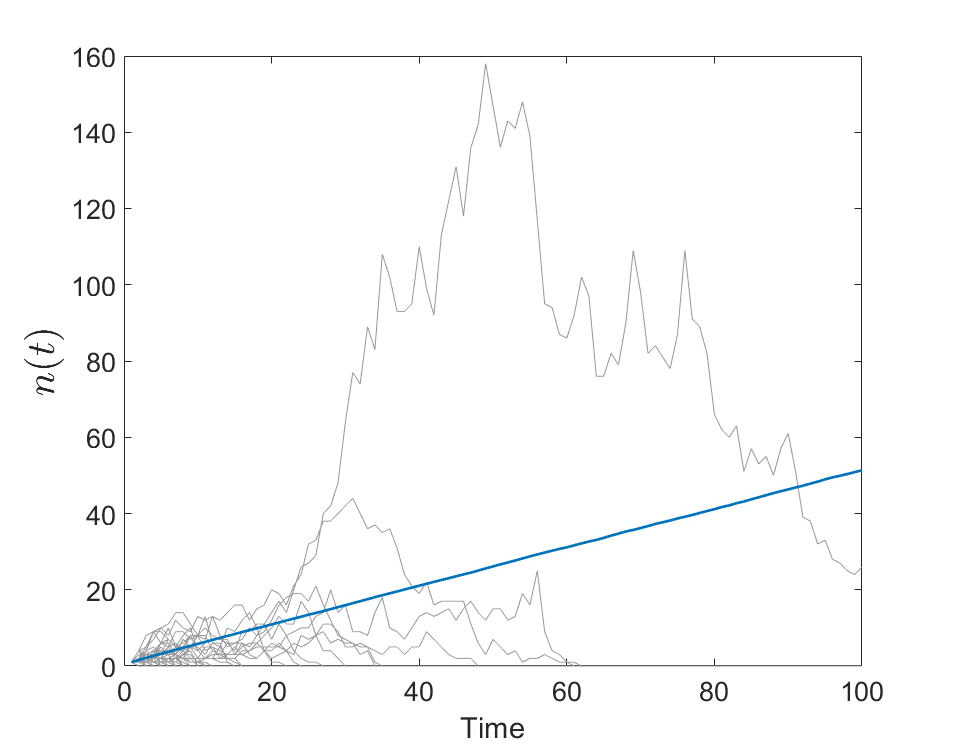} \includegraphics[width=0.4\textwidth]{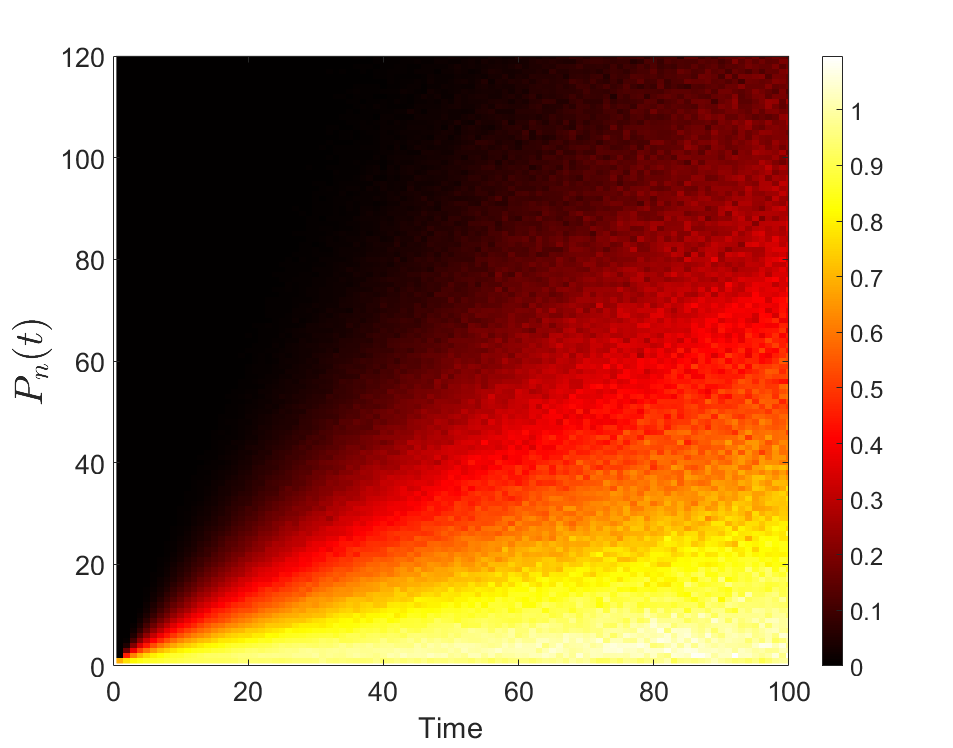} }
	\caption{\textbf{ \bf Growth of $n(t)$ with $t$ in a neutral process}. The left panel depicts several histories of $n_t$ as a function of $t$. in all cases, $n_0=1$, and $n_{t+1}$ was drawn from a Poisson distribution whose mean is $n_t$. Each gray line represents a single history. Most of these histories went extinct in the first steps, in agreement with Eq. (\ref{Eqap1}). The surviving histories reach higher population levels, and the mean (over the surviving populations) population size grows linearly with $t$ (blue line, average taken over $10^6$ histories). Here we implemented a Poisson process for which $\sigma^2 =1$, therefore the slope of the blue line is $1/2$. The right panel presents a heat map, where the color of $n_t$ corresponds to the number of visits at $n$ at time $t$, divided by the number of processes that survive until $t$ and multiplied by $t$.        \label{fig_ap1}}
\end{figure}

One may also approach the neutral process using the corresponding Fokker-Planck equation, 
\begin{equation}
\frac{dP(n)}{dt} = \frac{{\sigma^2}}{2} \frac{\partial }{\partial n} \left( nP(n)\right),
\end{equation}
where $\sigma^2$ is the variance of the number of offspring per generation. A simple scaling argument shows that for a general distribution of offspring number, as long as the process is neutral (i.e., its mean is unity), 
\begin{equation}
 {\overline n(t)} =\frac{\sigma^2 t}{2}.
 \end{equation}
 These features are demonstrated in Figure \ref{fig_ap1}.

\section{Simulation procedure} \label{apR}

\begin{figure*}[hbt!]
	\centering{
		\includegraphics[width=0.45\textwidth]{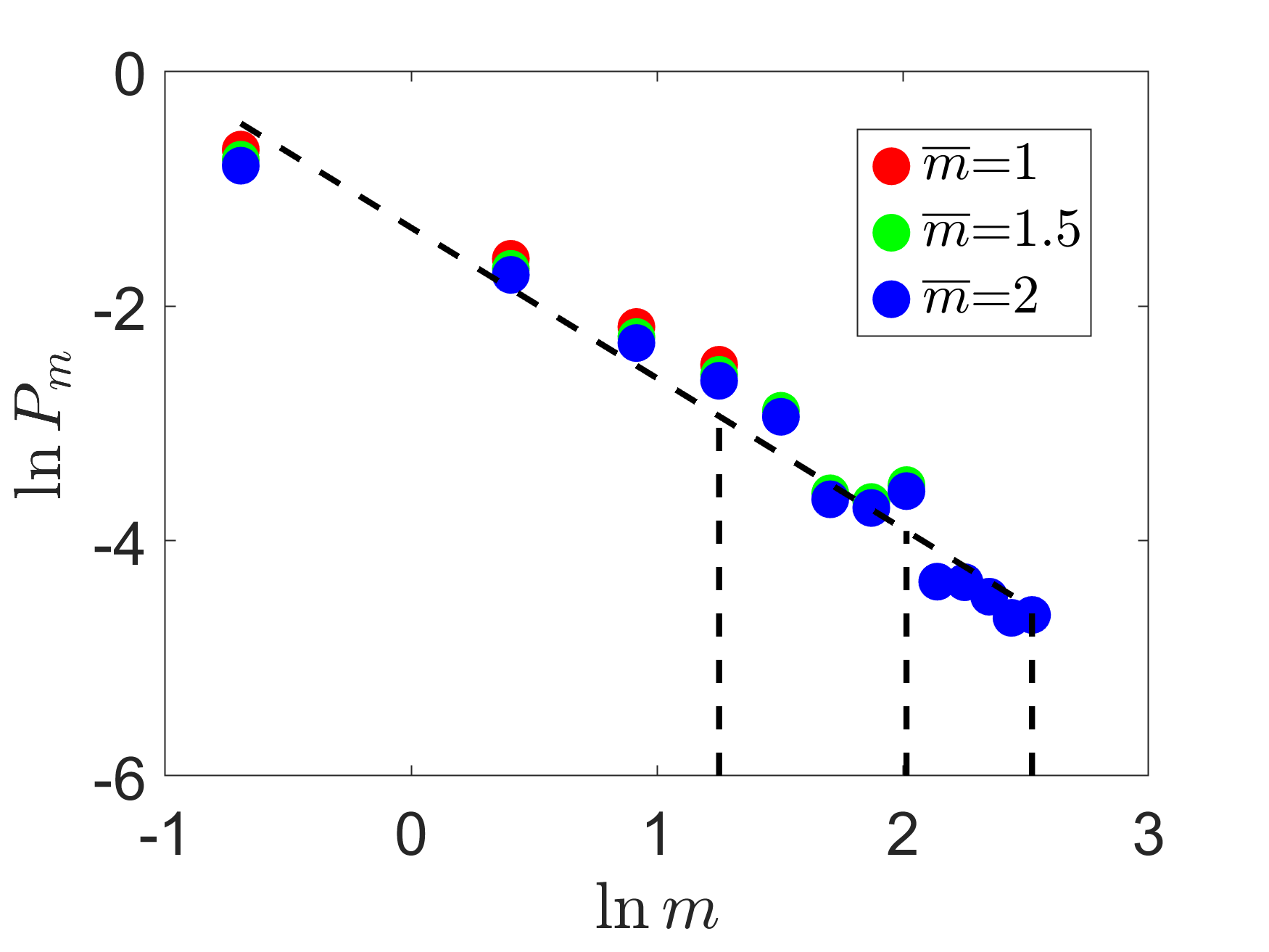}}
	\caption{\textbf{The probability distribution function \( P_m \)} for different values of \( \overline{m} \), plotted on a double logarithmic scale. \( P_m \) appears to follow a truncated power-law distribution, with an exponent of approximately \(-1.3\) (dashed line). To simulate the demographic process, the number of offspring per individual was drawn from this distribution, with the truncation point $m^*$ (vertical dashed lines) and normalization determined by the constraint on \( \overline{m} \).
   \label{figS31}}
\end{figure*}

To map the dynamics of interdependent networks to a birth--death process, we monitor the number of ``decedents'' per failed node, \( P_m \). When both networks are statistically identical, there is no need to track the process on interdependent networks. It is sufficient to follow an attack on a single network, where \( P_m \) is the probability that the removal of a single node detaches \( m \) other nodes from the giant connected component (GCC). Figure~\ref{figS31} shows the distribution of the number of offspring, \( P_m \) vs.\ \( m \), on a double logarithmic scale, for a few values of the mean number of offspring, \( \overline{m} \). As expected, this pdf is a truncated power law whose exponent is close to \( 1.3 \), and different values of \( \overline{m} \) differ only by the truncation point.

Hence, in the simulations of the demographic process (e.g., those shown in Figure~\ref{fig3} of the main text), we picked the number of offspring per individual from a truncated power law with a hard cutoff, i.e.,  \( P_m = A m^{-1.3} \) dor $m<m^*$ and $P_m=0$ for $m>m^*$,  where the truncation point $m^*$ and the normalization factor \( A \) were determined such that \( \overline{m} = 1 + CM/N \). The factor \( C \) depends on the slope of \( \overline{m} \) vs.\ \( M/N \) near the point where \( \overline{m} = 1 \), a slope that was very difficult to measure numerically. Therefore, we took the value \( C = 2.5 \) (which lies within the range of values we obtained) as representative, since it provided a better fit to the results.

\section{Additional details for Figure \ref{fig3}} \label{appE}

The results presented in Figure \ref{fig3} of the main text were obtained as follows. For every network size $N$, $10^4$ interdependent networks were generated. In each attack, $n_0$ sites were removed, and the resulting dynamics was monitored to determine $T_A$ and $T_F$. Six independent attacks with $n_0$ removed sites were performed for each of the $10^4$ realizations. The circles represent the mean, calculated over all these attacks.

\begin{figure*}[hbt!]
	\centering{
		\includegraphics[width=0.45\textwidth]{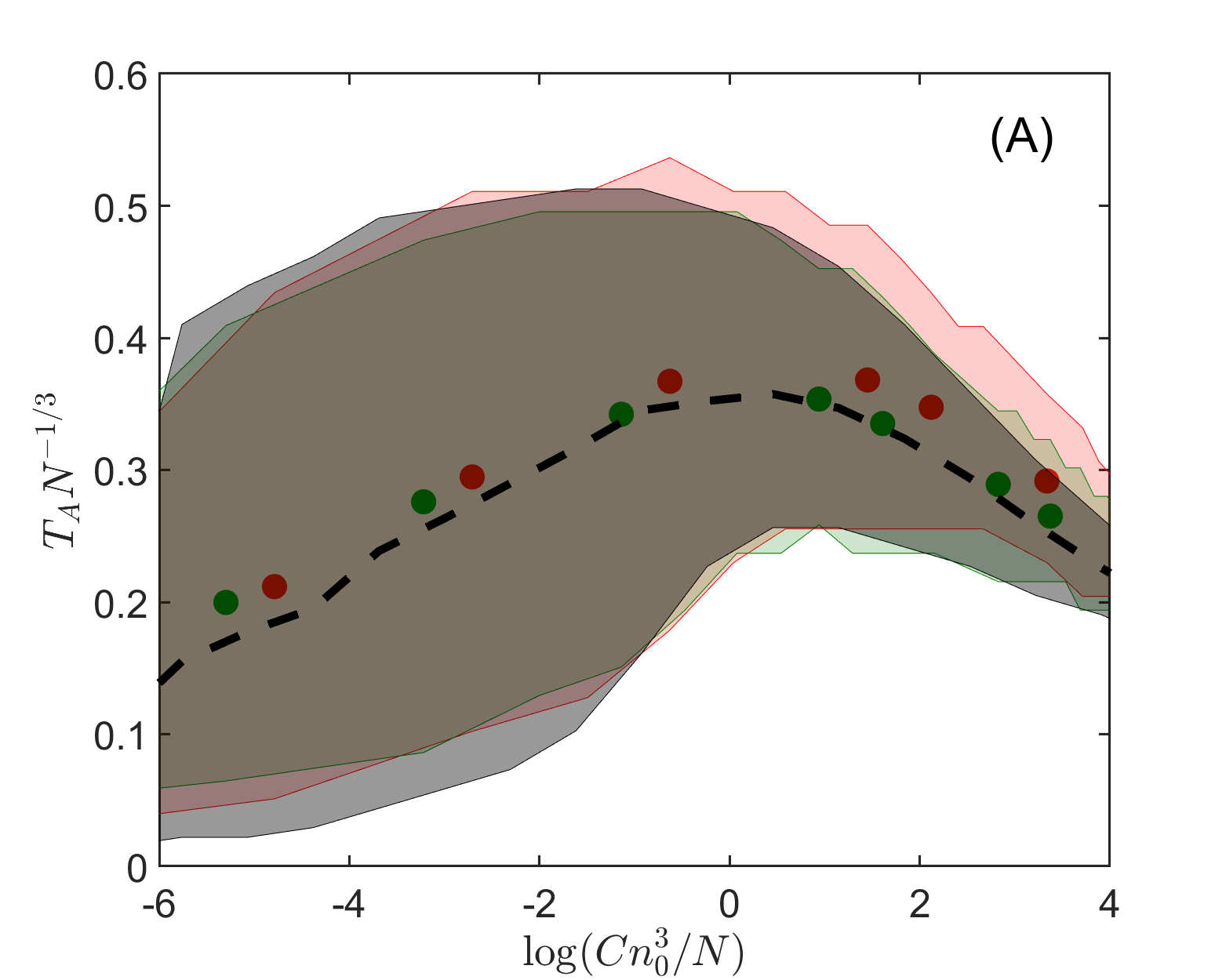}
\includegraphics[width=0.45\textwidth]{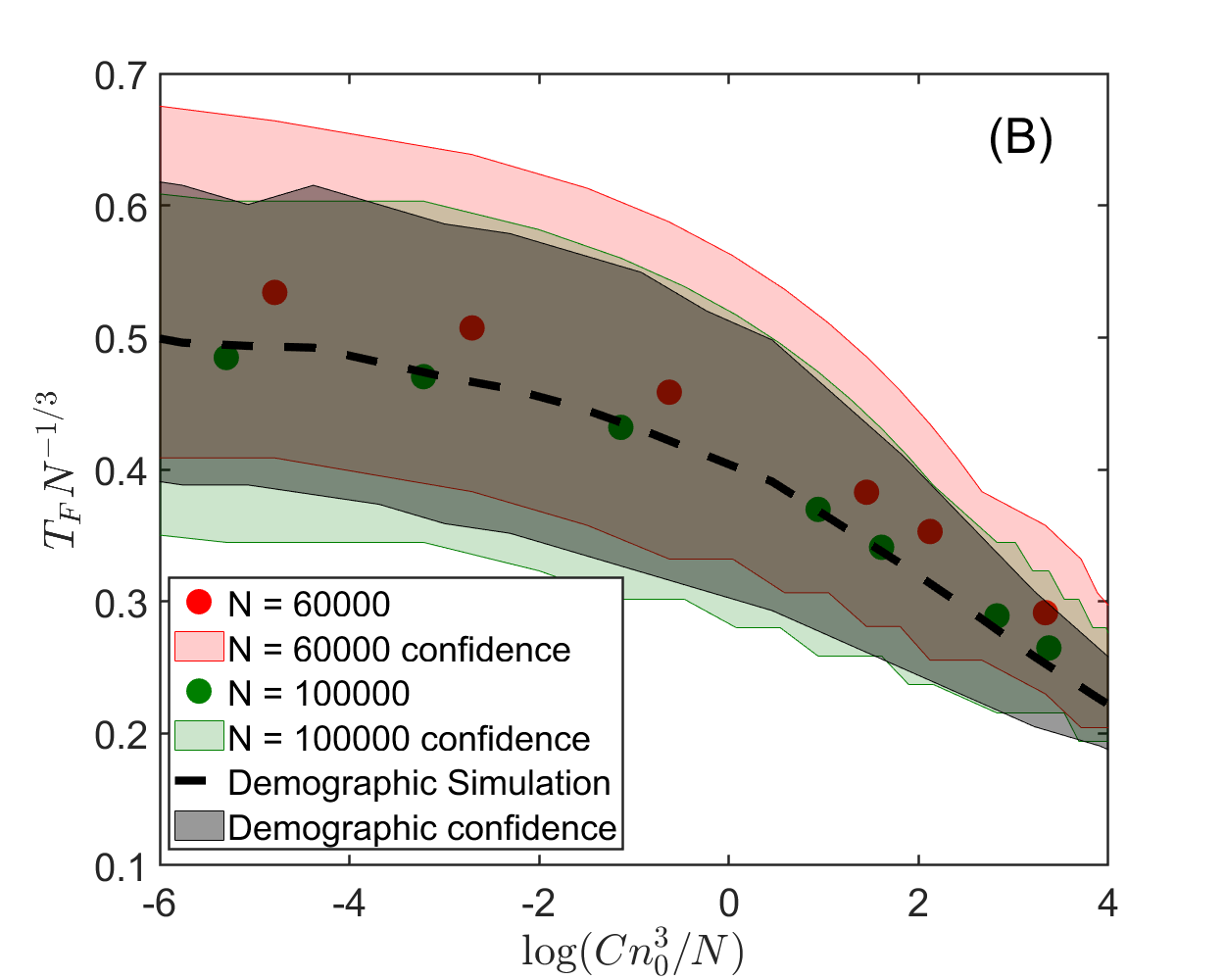}}
	\caption{{\bf Similar to Figure \ref{fig3} in the main text, with shaded areas indicating the width of the error bars.} Panel \textbf{(a)} shows the mean time to absorption $T_A$  and Panel \textbf{(b)}, the mean time to failure $T_F$. In $68\%$ (one standard deviation above or below the mean) of the simulations, the results fall within the shaded area. The results present over $10^4$ different networks for each $N$. \label{figS30}}
\end{figure*}

The spread of the results is shown in Figure \ref{figS30}, where the shaded areas mark the regions containing $67\%$ of the numerical experiment outcomes. Importantly, the demographic process accurately reproduces both the mean and the variance of the interdependent network dynamics. 

\end{document}